\documentclass[9pt,twocolumn,twoside]{arxiv}
\journal{arxiv} 
\setboolean{shortarticle}{false}

\usepackage{amsmath,graphicx}
\usepackage{bm}
\usepackage{mathtools} 
\usepackage{physics} 
\usepackage[textsize=tiny]{todonotes}
\usepackage{hyperref}
\usepackage[capitalize]{cleveref}
\usepackage{relsize}
\usepackage{soul,xcolor}
\usepackage[justification=justified]{caption}
\usepackage{lineno} 
\allowdisplaybreaks[4] 

\title{Single-molecule orientation localization microscopy I: fundamental limits}

\author[1]{Oumeng Zhang}
\author[1,2,3,*]{Matthew D. Lew}

\affil[1]{Department of Electrical and Systems Engineering, Washington University in St. Louis, MO 63130 USA}
\affil[2]{Center for the Science and Engineering of Living Systems, Washington University in St. Louis, MO 63130 USA}
\affil[3]{Institute of Materials Science and Engineering, Washington University in St. Louis, MO 63130 USA}
\affil[*]{mdlew@wustl.edu} 

\begin{abstract}
Precisely measuring the three-dimensional position and orientation of individual fluorophores is challenging due to the substantial photon shot noise in single-molecule experiments. Facing this limited photon budget, numerous techniques have been developed to encode 2D and 3D position and 2D and 3D orientation information into fluorescence images. In this work, we adapt classical and quantum estimation theory and propose a mathematical framework to derive the best possible precision for measuring the position and orientation of dipole-like emitters for any fixed imaging system. We find that it is impossible  to design an instrument that achieves the maximum sensitivity limit for measuring all possible rotational motions. Further, our vectorial dipole imaging model shows that the best quantum-limited localization precision is \textasciitilde4-8\% worse than that suggested by a scalar monopole model. Overall, we conclude that no single instrument can be optimized for maximum precision across all possible 2D and 3D localization and orientation measurement tasks.
\end{abstract}

\setboolean{displaycopyright}{true}

\begin{document}

\maketitle

\nolinenumbers

\section{Introduction}
\label{sec:intro}

Single fluorescent molecules are indispensable tools for studying nanoscale structures and dynamics within biological and material systems \cite{Triller2008,Kasai2011,Tachikawa2011,Gahlmann2014,lippert2017angular}, especially since the invention of super-resolution single-molecule localization microscopy (SMLM) nearly 15 years ago \cite{Hess2006,Betzig2006,Rust2006,Sharonov2006}. The field continues to innovate with interferometric \cite{Shtengel2009,Aquino2011,Huang2016} and adaptive \cite{Balzarotti2017,Gwosch2020,Cnossen2020} imaging techniques pushing practical localization precision to the molecular scale (\textasciitilde1~nm), much closer to the quantum limit in two and three dimensions \cite{backlund2018,tsang2019,prasad2019}. Beyond standard SMLM, microscopists have also developed many methods to measure the orientations and rotational diffusion, i.e. ``wobble,'' of single molecules (SMs) \cite{Backlund2014}. Recent developments in SM orientation localization microscopy (SMOLM) include new engineered point spread functions (PSFs) for measuring 3D orientation and wobble simultaneously \cite{zhang2018imaging,curcio2019birefringent,Lu2020}, the use of polarizers and/or phase masks to remove localization errors \cite{lew2014azimuthal,backlund2016removing,Nevskyi2020}, and leveraging supercritical light for improved sensitivity \cite{Ding2020}. 

These new techniques advance orientational measurement performance closer to the classical \cite{zhang2019fundamental} and quantum \cite{zhang2020quantum} sensitivity limits, typically quantified using the classical and quantum Cram\'er-Rao bounds (CRBs) \cite{moon2000mathematical}, respectively. Such quantitative analysis is essential, because fluorescence photons are precious and performance trade-offs are often encountered in multiparameter estimation \cite{zhang2020quantum}. Unfortunately, the forward imaging models \cite{Chandler:19,Chandler2019,Chandler2020} used to derive fundamental bounds for 3D localization \cite{backlund2018} and 3D orientation measurements \cite{zhang2020quantum} are not unified, thereby making it difficult to characterize and compare various approaches for SMOLM. From a design perspective, it is unknown how localization and orientation estimation sensitivities are coupled, and therefore, it is difficult to optimize an instrument for a particular orientation and localization task.

In this paper, we establish fundamental precision limits for measuring the rotational motions of single molecules that arise from the simple non-negativity property of photon counting. These limits necessitate a performance trade-off when designing SMOLM imaging systems; a single instrument cannot measure both the orientation and wobble of a molecule with maximum precision simultaneously. In the context of these limits, we compare the orientation measurement precisions of various imaging techniques. We then recapitulate and expand upon existing quantum estimation theory to derive the best-possible uncertainty for localizing a single molecule using a vectorial dipole emission model \cite{backer2014extending,backer2015determining,Stallinga2015,Chandler:19}. Critically, this study establishes the framework needed to compare various microscope configurations to fundamental performance limits, as well as optimize new SMOLM designs that balance fundamental performance trade-offs. An in-depth performance comparison of techniques for SMOLM is reserved for a second paper in this series \cite{location-orientation_2}.

\section{Image formation and Fisher information for measuring molecular orientation}
\label{sec:model}
\begin{figure*}[ht!]
    \centering
    \includegraphics[width=18.4cm]{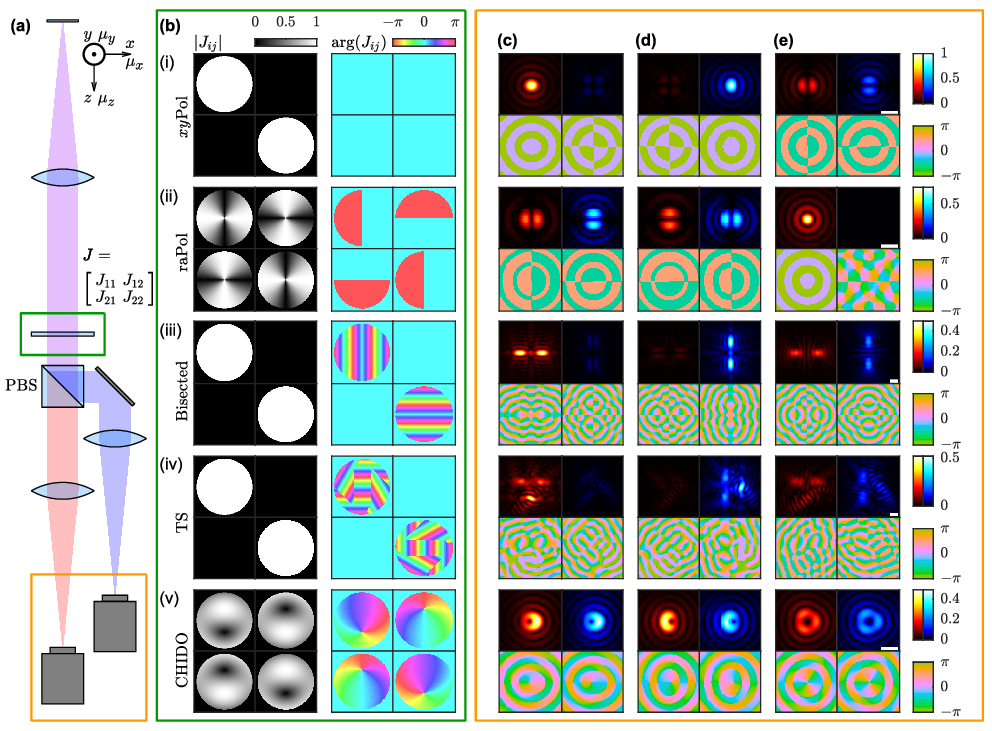}
    \caption{(a) Schematic of a polarization-sensitive $4f$ imaging system appended to the detection path of a fluorescence microscope. A polarizing beamsplitter (PBS) separates fluorescence photons (purple) into $x$ (red) and $y$ (blue) polarized detection channels. (b)~An optical component represented by a $2\times2$ complex polarization tensor $\bm{J}(u,v)$ is placed at the back focal plane (BFP, green). The amplitude $|J_{ij}(u,v)|$ and phase $\text{arg}(J_{ij}(u,v))$ correspond to polarization rotation and phase modulation, respectively, needed to implement the (i) $x$- and $y$-polarized ($xy$Pol) standard PSF, (ii) radially and azimuthally polarized (raPol) standard PSF , (iii) Tri-spot (TS) PSF, (iv) Bisected PSF, and (v) CHIDO (with a stress coefficient of $1.2\pi$). We assume a 90-degree rotation of the (iii) bisected and (iv) TS phase masks between $x$- and $y$-polarized imaging channels, which maximizes measurement precision. Tube lenses in front of each detector perform a Fourier transform on the modulated optical field at the BFP, yielding the basis fields (c) ${G}_x(\xi,\eta)$, (d) ${G}_y(\xi,\eta)$, and (e) ${G}_z(\xi,\eta)$ at the image plane (orange). Top row:  amplitude (normalized to the $xy$Pol standard PSF), bottom row: phase (rad). Scale bar: 500 nm.}
    \label{fig:1}
\end{figure*}

We model fluorescent molecules as oscillating electric dipoles positioned at $\bm{r}(t)=[x(t),y(t),z(t)]^\intercal$ with an orientation unit vector $\bm{\mu}(t)=[\mu_x(t),\mu_y(t),\mu_z(t)]^\intercal$ at time $t$ [\cref{fig:1} (a)], where we use $[\cdot]^\intercal$ to denote the transpose of a vector or matrix. We assume photons are emitted spontaneously by the dipole, are quasimonochromatic, and have wavelength $\lambda$ in the imaging medium. Once the collected photons travel to the pupil of an ideal objective lens, we assume they undergo paraxial propagation through lossless linear optical components to an image plane. The photons are absorbed by a photon-counting detector over a non-zero integration time. The electric field $E_0^{(l)}$ at the back focal plane (BFP), corresponding to the fluorescence light captured by an objective lens, is given by \cite{bohmer2003orientation,lieb2004single,axelrod2012fluorescence,novotny2012principles,backer2014extending,backer2015determining,Chandler:19}
\begin{linenomath}\begin{align}
\label{eqn:model}
    \bm{E}_0(u,v)=\ &\left[E_{0}^{(x)},E_{0}^{(y)},E_{0}^{(z)}\right]^\intercal \nonumber\\
    =\ &\exp\left\{j\,k\left[x(t)u+y(t)v+z(t)\sqrt{1-u^2-v^2}\right]\right\} \nonumber\\
    &\begin{bmatrix}g_{x}^{(x)}(u,v)&g_{y}^{(x)}(u,v)&g_{z}^{(x)}(u,v)\\
    g_{x}^{(y)}(u,v)&g_{y}^{(y)}(u,v)&g_{z}^{(y)}(u,v)\\
    0&0&0\end{bmatrix}
    \begin{bmatrix}\mu_x(t)\\\mu_y(t)\\\mu_z(t)\end{bmatrix},
\end{align}\end{linenomath}
where $g_i^{(l)}$ are the basis fields in linearly $l$-polarized detection channels produced by an in-focus dipole with orientation $\mu_i(t)$ as defined previously \cite{novotny2012principles,backer2014extending,backer2015determining,zhang2020quantum}, and $\{i,l\} \in \{x,y,z\}$. The wavenumber $k$ is given by $2\pi/\lambda$, and the BFP coordinates $(u,v)$ are normalized such that $u^2+v^2<1$. Further, when the refractive index (RI) of the sample $n$ matches that of the lens immersion medium, $g_i^{(l)}(u,v)=0$ for $u^2+v^2>(\text{NA}/n)^2$.
For simplicity, we define the vectorial basis field emitted by a dipole with orientation $\mu_i(t)$ as
\begin{linenomath}
\begin{subequations} \label{eqn:basis_field}
\begin{align}
    \check{\bm{g}}_i(u,v;x,y,z) &= \exp\left[j\,\theta(u,v;x,y,z)\right]
    \begin{bmatrix}{g}_i^{(x)}\\{g}_i^{(y)}\end{bmatrix}, \text{ where} \\
    \theta(u,v;x,y,z) &= k\left(xu+yv+z\sqrt{1-u^2-v^2}\right) \label{eqn:phase_delay}
\end{align}
\end{subequations}
\end{linenomath}
is the phase delay that arises from the emitter's position $[x,y,z]^\intercal$.

In this work, we focus on translationally fixed molecules, i.e., $\bm{r}(t)=\bm{r}=[x,y,z]^\intercal$. The resulting electric field in the image plane from a molecule with orientation $\bm{\mu}(t)$ is thus given by
\begin{linenomath}\begin{align}
    {E}(\xi,&\eta;\,\bm{\mu}(t),\bm{r})\nonumber =
    E^{(x)}(\xi_x,\eta_x;\,\bm{\mu}(t),\bm{r})\oplus E^{(y)}(\xi_y,\eta_y;\,\bm{\mu}(t),\bm{r})\nonumber\\
    =\ & U\left(\check{\bm{g}}_x(u,v;\,\bm{r})\mu_x(t)+\check{\bm{g}}_y(u,v;\,\bm{r})\mu_y(t)+\check{\bm{g}}_z(u,v;\,\bm{r})\mu_z(t)\right) \nonumber\\
    =\ & U\left(\check{\bm{g}}_x(u,v;\,\bm{r})\right)\mu_x(t)+U\left(\check{\bm{g}}_y(u,v;\,\bm{r})\right)\mu_y(t)\nonumber \\
    &+U\left(\check{\bm{g}}_z(u,v;\,\bm{r})\right)\mu_z(t) \nonumber\\
    =\ &{G}_x(\xi,\eta;\,\bm{r})\mu_x(t)+{G}_y(\xi,\eta;\,\bm{r})\mu_y(t)+{G}_z(\xi,\eta;\,\bm{r})\mu_z(t), \label{eqn:basisFieldsImagePlane}
\end{align} \end{linenomath}
where $E^{(x)}(\xi_x,\eta_x)\oplus E^{(y)}(\xi_y,\eta_y)$ represents a non-overlapping spatial sum of $x$- and $y$-polarized fields, i.e., the two polarizations are detected separately and simultaneously. For convenience, we define a unified coordinate system $[\xi,\eta]\in\mathbb{R}^2_{(x)}\cup\mathbb{R}^2_{(y)}$ such that $\mathbb{R}^2_{(x)}\cap\mathbb{R}^2_{(y)}=\varnothing$, where $[\xi_x,\eta_x]\in\mathbb{R}^2_{(x)}$ and $[\xi_y,\eta_y]\in\mathbb{R}^2_{(y)}$. The image plane electric field $E$ is thus given by
\begin{equation}
    E(\xi,\eta)=\left\{
    \begin{matrix}
    E^{(x)}(\xi,\eta) & \text{if }[\xi,\eta]\in\mathbb{R}^2_{(x)}\\
    E^{(y)}(\xi,\eta) & \text{if }[\xi,\eta]\in\mathbb{R}^2_{(y)}\end{matrix}
    \right. .
    \label{eqn:nonOverlappingSum}
\end{equation}
The unitary operator $U(\cdot)$ models the (linear) imaging system, and ${G}_i(\xi,\eta;\,\bm{r})$ represents the time-invariant basis field sampled in the image plane $(\xi,\eta)$ corresponding to orientation component $\mu_i$. For a typical imaging system [\cref{fig:1}(a)], $U(\cdot)$ models polarization and/or phase modulation of light at the BFP followed by a Fourier transform performed by a tube lens, i.e.,
\begin{linenomath}\begin{subequations}
\begin{align}
    {G}_i(\xi,\eta)=U\left(\bm{g}_i(u,v)\right)&=G_i^{(x)}(\xi_x,\eta_x)\oplus G_i^{(y)}(\xi_y,\eta_y) \text{ and}\\
    \begin{bmatrix}G_i^{(x)}(\xi_x,\eta_x)\\G_i^{(y)}(\xi_y,\eta_y)\end{bmatrix}
    &=
    \mathcal{F}\left\{\bm{J}(u,v)\begin{bmatrix}g_i^{(x)}(u,v)\\g_i^{(y)}(u,v)\end{bmatrix}\right\},
\end{align}
\end{subequations}\end{linenomath}
where a Jones matrix $\bm{J}(u,v)$ represents a spatially-varying $2\times2$ polarization tensor \cite{saleh2019fundamentals} and $\mathcal{F}$ represents a spatial 2D Fourier transform \cite{goodman2005introduction}.

To graphically illustrate and compare various imaging techniques, we show corresponding polarization tensors $\bm{J}(u,v)$ [\cref{fig:1}(b)] and basis fields ${G}_i(\xi,\eta)$ [\cref{fig:1}(c-e)] for in-focus molecules, i.e., $z=0$. For the $x$- and $y$-polarized standard PSF [$xy$Pol, \cref{fig:1}(i), \cite{mortensen2010optimized}], light is unmodulated at the BFP; therefore $\bm{J}$ is an identity matrix. To model spatially varying polarization modulation, the tensor $\bm{J}$ is a product of a spatially varying real tensor and a spatially uniform phase mask. One example is a vortex (half) wave plate (VWP) placed in the BFP to convert radially and azimuthally polarized light to $x$- and $y$-polarized light; i.e., the camera resolves radially  and azimuthally polarized light emitted by the molecule separately [raPol, \cref{fig:1}(ii), \cite{lew2014azimuthal,backlund2016removing}]. For phase modulation, e.g., the Bisected [\cref{fig:1}(iii), \cite{backer2014bisected}] and Tri-spot [TS, \cref{fig:1}(iv), \cite{zhang2018imaging}] PSFs, the off-diagonal polarization mixture terms in $\bm{J}$ vanish, i.e., $J_{12}=J_{21}=0$. Similar to $xy$Pol, the detected light for $x$- and $y$-oriented molecules are concentrated in their respective $x$- and $y$-polarization channels in these techniques. Newly developed methods, e.g., Coordinate and Height super-resolution Imaging with Dithering and Orientation (CHIDO) [\cref{fig:1}(v), \cite{curcio2019birefringent}], combine phase and polarization modulations using stressed-engineered optics (SEO) to create entirely unique basis fields ${G}_i$.

Fundamentally, cameras are sensitive to intensity $I=|E|^2$, not electric field. Each detected photon arises from a specific molecular orientation $\bm{\mu}(t)$ at the instant it is emitted. Thus, the captured intensity distribution can be written as a temporal average over the acquisition time $T$, given by
\begin{linenomath}\begin{align}
\label{eqn:foward_model}
    I(\xi,\eta;\bm{m}) =\ &\frac{1}{T}\int_0^T \left|{E}\left(\xi,\eta;\bm{\mu}(t)\right)\right|^2 dt \nonumber \\
    =\ &B_{xx}m_{xx}+B_{yy}m_{yy}+B_{zz}m_{zz} \nonumber\\
    &+B_{xy}m_{xy}+B_{xz}m_{xz}+B_{yz}m_{yz},
\end{align}\end{linenomath}
where
\begin{equation}
    m_{ij} = \frac{1}{T}\int_0^T\mu_i(t)\mu_j(t)dt
\end{equation}
are the second-order orientational moments, $\{i,j\} \in \{x,y,z\}$, and
\begin{linenomath}\begin{subequations}
\begin{align}
    B_{xx}(\xi,\eta) &= \left| G_x(\xi,\eta) \right| ^2 \\
    B_{yy}(\xi,\eta) &= \left| G_y(\xi,\eta) \right| ^2 \\
    B_{zz}(\xi,\eta) &= \left| G_z(\xi,\eta) \right| ^2 \\
    B_{xy}(\xi,\eta) &= G_x(\xi,\eta) G_y^*(\xi,\eta) + G_x^*(\xi,\eta) G_y(\xi,\eta) \\
    B_{xz}(\xi,\eta) &= G_x(\xi,\eta) G_z^*(\xi,\eta) + G_x^*(\xi,\eta) G_z(\xi,\eta) \\
    B_{yz}(\xi,\eta) &= G_y(\xi,\eta) G_z^*(\xi,\eta) + G_y^*(\xi,\eta) G_z(\xi,\eta)
\end{align}
\end{subequations}\end{linenomath}
are the basis images ${B}_{ij}(\xi,\eta)$ that correspond to each second moment $m_{ij}$. Here, $(\cdot)^*$ represents complex conjugation. To facilitate quantitative comparisons without loss of generality, we normalize the summed intensity contained within these basis images as
\begin{linenomath}\begin{subequations} 
\label{eqn:basis_normalization}
\begin{align}
    \iint B_{xx}\,d\xi\,d\eta &= \iint B_{yy}\,d\xi\,d\eta = 1 \text{ and}\\
    \iint B_{zz}\,d\xi\,d\eta &= c =\frac{4-2\epsilon-2\epsilon^2}{4+\epsilon+\epsilon^2} \le 1. \label{eqn:normFac}
\end{align}
\end{subequations}\end{linenomath}
The constant $\epsilon = \sqrt{1-(\text{NA}/n)^2}$ accounts for the relative inefficiency of photon collection from $z$-oriented dipoles \cite{zhang2020quantum}. Note that the second equality in \cref{eqn:normFac} is only satisfied if the sample RI matches that of the lens immersion medium; the normalized total intensity $c$ contained within the $B_{zz}$ basis image may be greater than that contained within $B_{xx}$ or $B_{yy}$ if supercritical light is collected \cite{location-orientation_2}.

Given the forward imaging model in \cref{eqn:foward_model}, measuring molecular orientation can be viewed as estimating the time-averaged second-order moments $\bm{m}$ of $\bm{\mu}(t)$ given a captured image $\hat{I}(\xi,\eta)$. To quantitatively evaluate the performance of any imaging system $U(\cdot)$ for measuring the second moments $m_{ij}$, we use the Cram\'er-Rao bound (CRB) $\bm{\mathcal{J}_m}^{-1}$, which bounds the best-possible variance $\bm{V_m}$ achievable by any unbiased estimator such that $\bm{V_m}-\bm{\mathcal{J}_m}^{-1}$ is always positive semidefinite. The $6\times6$ Fisher information (FI) matrix $\bm{\mathcal{J}_m}$ is given by
\begin{equation}
    \bm{\mathcal{J}_m}= \begin{bmatrix} \mathcal{J}_{xx,xx} & \cdots & \mathcal{J}_{xx,yz} \\ \vdots &\ddots &\vdots\\
    \mathcal{J}_{yz,xx}& \cdots & \mathcal{J}_{yz,yz}
    \end{bmatrix} =
    \begin{bmatrix} \mathcal{J}_{11} & \cdots & \mathcal{J}_{16} \\ \vdots &\ddots &\vdots\\
    \mathcal{J}_{61}& \cdots & \mathcal{J}_{66}
    \end{bmatrix}.
\end{equation}
The entries of $\bm{\mathcal{J}_m}$ measure the changes of the image with respect to the corresponding parameters of interest, given by
\begin{equation}
    \mathcal{J}_{ij,kl} = \iint \frac{{B}_{ij}(\xi,\eta) {B}_{kl}(\xi,\eta)}{I(\xi,\eta)}\,du\,dv,
\end{equation}
where combinations of subscripts $\{i,j,k,l\} \in \{x,y,z\}$ denote the six orientational second moments.

Although quantitatively calculating $\bm{\mathcal{J}_m}$ for an imaging system often requires numerical simulations, some qualitative properties of certain entries of $\bm{\mathcal{J}_m}$ can be predicted by examining the basis fields [\cref{eqn:basisFieldsImagePlane,fig:1}(c-e)]. For example, basis fields ${G}_z$ corresponding to both $xy$Pol [\cref{fig:1}(e)(i)] and raPol [\cref{fig:1}(e)(ii)] standard PSFs are shifted in phase by $\pi/2$ relative to ${G}_x$ [\cref{fig:1}(c)(i,ii)] and ${G}_y$ [\cref{fig:1}(d)(i,ii)] at all positions in the image plane $(\xi,\eta)$. Therefore, these fields will not interfere with each other; i.e., ${B}_{xz}={B}_{yz}=0$ and all corresponding FI entries are zero. This lack of sensitivity is typically caused by parity symmetry in the tensor $\bm{J}$ when all entries are real, i.e., for polarization-sensitive imaging systems without phase masks and negligible optical aberrations.

For imaging systems that separate $x$- and $y$-polarized light and use phase masks, such as the TS and bisected PSFs, the energy of ${G}_x$ [\cref{fig:1}(c)(iii,iv)] and ${G}_y$ [\cref{fig:1}(d)(iii,iv)] are separated into different detection channels, and thus ${B}_{xy}$ is much weaker than the other basis images. Therefore, the corresponding FI entry $\mathcal{J}_{xy,xy}=\mathcal{J}_{44}$ is often much smaller than other diagonal entries. For example, for the $xy$Pol system imaging a freely-rotating molecule with one photon detected, $\mathcal{J}_{11}=\mathcal{J}_{22}$ is \textasciitilde10 times stronger and $\mathcal{J}_{33}$ is \textasciitilde4 times stronger than $\mathcal{J}_{44}$.

\section{Optimal methods for estimating the orientation of fluorescent molecules}
\label{sec:orientation}
The goal of PSF engineering is to optimize the imaging system operator $U$ or, more specifically, the tensor $\bm{J}$ and therefore improve the measurement precision of an imaging system. However, joint optimization over all six orientational second moments is computationally expensive, and previous analyses have shown that it is not possible to achieve the best-possible measurement precision across all six moments simultaneously \cite{zhang2020quantum}.
Instead of searching for an optimal imaging system tensor $\bm{J}$, we directly derive an upper bound on the FI matrix itself using physical constraints, i.e., photon non-negativity; an imaging system can be considered optimal if it performs closely to this bound.

Since the determinant of the covariance matrix is often used to evaluate the scatter of multidimensional estimates around their mean, we use the $[-1/(2N)]^\text{th}$ power of the determinant of the $N\times N$ FI matrix, i.e., the square root of the standardized generalized variance (SGV) \cite{sengupta1987tests}, as a metric to measure the overall precision of measuring the second-order orientational moments. This quantity can be viewed as the geometric average of the standard deviations of all parameters in the measurement. Other performance metrics, e.g., an arithmetic mean of the CRB across all second moments, can also be evaluated in a similar manner.

The determinant of the FI matrix is a function of all its entries, i.e., 21 independent values for the 6 orientational second-order moments. Therefore, it is still difficult to directly find a bound. Here, we apply some constraints to reduce the number of independent FI entries and simplify the computation. Firstly, we remove some entries that can be bounded by zero in the FI matrix, i.e.,
\begin{linenomath}\begin{multline}
    \det{\bm{\mathcal{J}}}\leq
    \det\{ \bm{\mathcal{J}}_{xx,yy,zz} \} \det\{\bm{\mathcal{J}}_{xy,xz,yz}\} \\
    \leq \det{ \bm{\mathcal{J}}_{xx,yy,zz} } \mathcal{J}_{44}\mathcal{J}_{55}\mathcal{J}_{66},
\end{multline}\end{linenomath}
where
\begin{linenomath}\begin{subequations}
\begin{align}
    \bm{\mathcal{J}}_{xx,yy,zz} &= 
        \begin{bmatrix} \mathcal{J}_{11}&\mathcal{J}_{12}&\mathcal{J}_{13}\\
        \mathcal{J}_{12}&\mathcal{J}_{22}&\mathcal{J}_{23}\\
        \mathcal{J}_{13}&\mathcal{J}_{23}&\mathcal{J}_{33} \end{bmatrix} \text{ and}\\
    \bm{\mathcal{J}}_{xy,xz,yz} &= 
        \begin{bmatrix} \mathcal{J}_{44}&\mathcal{J}_{45}&\mathcal{J}_{46}\\
        \mathcal{J}_{45}&\mathcal{J}_{55}&\mathcal{J}_{56}\\
        \mathcal{J}_{46}&\mathcal{J}_{56}&\mathcal{J}_{66} \end{bmatrix},
\end{align}
\end{subequations}\end{linenomath}
since the determinants of positive definite matrices are bounded by the product of their diagonal entries (Hadamard's inequality). The equalities are satisfied when $\mathcal{J}_{ij}=0$ for $j\in\{4,5,6\}$, $i\in\{1,2,3,4,5,6\}$, and $i\neq j$.

Next, we evaluate the precision of estimating the cross moments $m_{xy}$, $m_{xz}$ and $m_{yz}$. As shown in \cite{zhang2020quantum}, the brightness of the basis images $B_{xy}$, $B_{xz}$ and $B_{yz}$ are bounded as 
\begin{linenomath}\begin{subequations}
\begin{align}
    B_{xy}^2(\xi,\eta)&\leq4B_{xx}(\xi,\eta)B_{yy}(\xi,\eta),\\
    B_{xz}^2(\xi,\eta)&\leq4B_{xx}(\xi,\eta)B_{zz}(\xi,\eta), \text{ and}\\
    B_{yz}^2(\xi,\eta)&\leq4B_{yy}(\xi,\eta)B_{zz}(\xi,\eta),
\end{align}
\end{subequations}\end{linenomath}
with equalities if and only if $G_x$, $G_y$ and $G_z$ have the same phase.
Therefore, the corresponding FI entries are also bounded as
\begin{linenomath}\begin{subequations}
\begin{align}
    \mathcal{J}_{44}&=\iint\frac{B_{xy}^2}{I}\,d\xi\,d\eta\leq\iint\frac{4B_{xx}B_{yy}}{I}\,d\xi\,d\eta = 4\mathcal{J}_{12},\\
    \mathcal{J}_{55}&=\iint\frac{B_{xz}^2}{I}\,d\xi\,d\eta\leq\iint\frac{4B_{xx}B_{zz}}{I}\,d\xi\,d\eta = 4\mathcal{J}_{13}, \text{ and}\\
    \mathcal{J}_{66}&=\iint\frac{B_{yz}^2}{I}\,d\xi\,d\eta\leq\iint\frac{4B_{yy}B_{zz}}{I}\,d\xi\,d\eta = 4\mathcal{J}_{23}.
\end{align}
\end{subequations}\end{linenomath}
Thus, we can bound the maximum determinant of the FI matrix from above, and thus the maximum sensitivity of any imaging system, by defining a cost function $f:\mathbb{R}^6\xrightarrow{}\mathbb{R}$ as
\begin{equation}
    f(\bm{\mathcal{J}})=\left(\det{\bm{\mathcal{J}}_{xx,yy,zz}}\mathcal{J}_{12}\mathcal{J}_{13}\mathcal{J}_{23}\right)^{-1}, \label{eqn:optimizationCost}
\end{equation}
where $\det\{\bm{\mathcal{J}}\}\leq (f\{\bm{\mathcal{J}}\})^{-1}$. Minimizing this cost function $f(\bm{\mathcal{J}})$ is equivalent to minimizing the SGV, or total uncertainty, for measuring the second-order orientational moments $\bm{m}$.

To further simplify the function $f$, we assume a simplified model of rotational diffusion where a molecule symmetrically wobbles around an average orientation $\bar{\bm{\mu}}=[\bar{\mu}_x,\bar{\mu}_y,\bar{\mu}_z]^\intercal$ with a rotational constraint $\gamma\in[0,1]$ \cite{zhang2018imaging,zhang2019fundamental,curcio2019birefringent}. The second-order orientational moments are therefore given by 
\begin{linenomath}
\begin{subequations} \label{eqn:rotationSimpleModel}
\begin{align}
    m_{xx}&=\gamma\bar{\mu}_x^2+(1-\gamma)/3 \\ m_{yy}&=\gamma\bar{\mu}_y^2+(1-\gamma)/3 \\ m_{zz}&=\gamma\bar{\mu}_z^2+(1-\gamma)/3 \\ m_{xy}&=\gamma\bar{\mu}_x\bar{\mu}_y \\ m_{xz}&=\gamma\bar{\mu}_x\bar{\mu}_z \\ m_{yz}&=\gamma\bar{\mu}_y\bar{\mu}_z,
\end{align}
\end{subequations}
\end{linenomath}
where $\gamma=1$ represents a fixed molecule and $\gamma=0$ represents a freely rotating molecule. For additional simplicity, we consider molecules wobbling around a Cartesian axis, e.g., the $\mu_x$ axis, for which $\bar{\mu}_x=1$. Since the summed intensity with the basis images is normalized [\cref{eqn:basis_normalization}], the FI entry corresponding to $m_{xx}$ can be written as a function of off-diagonal elements $\mathcal{J}_{12}$ and $\mathcal{J}_{13}$, i.e., 
\begin{linenomath}\begin{align}
\label{eqn:FIconstraint_x1}
    &\frac{3}{1+2\gamma} - \mathcal{J}_{11}\nonumber\\
    =\ &\iint\frac{3B_{xx}^2\,d\xi\,d\eta}{(1+2\gamma)B_{xx}} - \iint \frac{3B_{xx}^2\,d\xi\,d\eta}{(1+2\gamma)B_{xx}+(1-\gamma)(B_{yy}+B_{zz})} \nonumber\\
    =\ &\iint\frac{3(1-\gamma)(B_{xx}B_{yy}+B_{xx}B_{zz})}{(1+2\gamma)[(1+2\gamma)B_{xx}+(1-\gamma)(B_{yy}+B_{zz})]}\,d\xi\,d\eta \nonumber\\
    =\ & \frac{1-\gamma}{1+2\gamma}(\mathcal{J}_{12}+\mathcal{J}_{13}),
\end{align}\end{linenomath}
regardless of the imaging system $U$ and basis images $\{B_{ij}\}$. Similarly, the FI entries corresponding to $m_{yy}$ and $m_{zz}$ also satisfy
\begin{linenomath}\begin{subequations}
\label{eqn:FIconstraint_x2}
\begin{align}
    \frac{3}{1-\gamma}-\mathcal{J}_{22} &= \frac{1+2\gamma}{1-\gamma}\mathcal{J}_{12}+\mathcal{J}_{23} \text{ and} \\ \frac{3c}{1-\gamma}-\mathcal{J}_{33} &= \frac{1+2\gamma}{1-\gamma}\mathcal{J}_{13} + \mathcal{J}_{23}.
    \end{align}
\end{subequations}\end{linenomath}
Therefore, for molecules wobbling around the $\mu_x$ axis, the optimal FI matrix is obtained when function $f$ is minimized under the constraints in \cref{eqn:FIconstraint_x1,eqn:FIconstraint_x2}, yielding an optimization over 3 dimensions: $(\mathcal{J}_{11}, \mathcal{J}_{22}, \mathcal{J}_{33})$. Similarly, for molecules wobbling around the $\mu_z$ axis, i.e., $\bar{\mu}_z=1$, the optimal FI matrix is obtained when $f$ is minimized under the constraints
\begin{linenomath}\begin{subequations}
\label{eqn:FIconstraint_z}
\begin{align}
    \frac{3}{1-\gamma}-\mathcal{J}_{11} &= \mathcal{J}_{12}+\frac{1+2\gamma}{1-\gamma}\mathcal{J}_{13}, \\
    \frac{3}{1-\gamma}-\mathcal{J}_{22} &= \mathcal{J}_{12}+\frac{1+2\gamma}{1-\gamma}\mathcal{J}_{23}, \text{ and} \\ \frac{3c}{1+2\gamma}-\mathcal{J}_{33} &= \frac{1-\gamma}{1+2\gamma}(\mathcal{J}_{13} + \mathcal{J}_{23}).
    \end{align}
\end{subequations}\end{linenomath}
\begin{figure}[ht!]
    \centering
    \includegraphics[width=8.8cm]{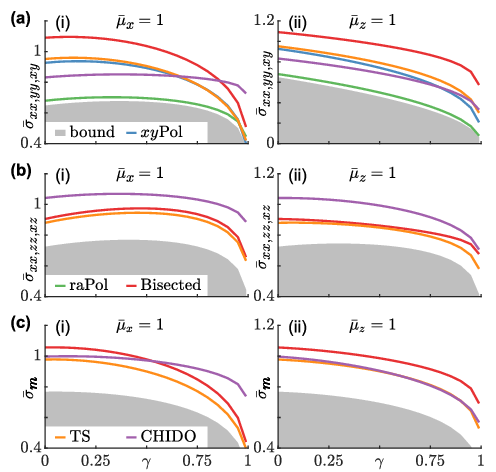}
    \caption{Limit of estimation precision (a) $\bar{\sigma}_{xx,yy,xy}$ for $m_{xx}$, $m_{yy}$, and $m_{xy}$; (b) $\bar{\sigma}_{xx,zz,xz}$ for $m_{xx}$, $m_{zz}$, and $m_{xz}$; and (c) $\bar{\sigma}_{\bm{m}}$ for all second-order orientational moments $\bm{m}$ for molecules wobbling around (i) the $\mu_x$ axis and (ii) the $\mu_z$ axis. The precision is calculated assuming one photon is detected; for $N$ photons detected, scale the values by $1/\sqrt{N}$. All techniques are compared to the best-possible standard generalized variance (SGV) of the CRB, which bounds the gray area from above using (a) \cref{eqn:xyBound}, (b)~\cref{eqn:xzBound}, and (c) numerical calculations [\cref{eqn:fullMbound}]. Blue: $xy$Pol standard PSF, green: raPol standard PSF, red: Bisected PSF, orange: TS PSF, purple: CHIDO.}
    \label{fig:2} 
\end{figure}

In the limiting case where $\mathcal{J}_{xz,xz}=\mathcal{J}_{yz,yz}=0$, we find an analytical bound on the combined precision of measuring second-order moments $m_{xx}$, $m_{yy}$, and $m_{xy}$ [\cref{fig:2}(a)], given by
\begin{linenomath}\begin{subequations}
\label{eqn:xyBound}
\begin{align}
    \left.\bar{\sigma}^{2}_{xx,yy,xy}\right|_{\bar{\mu}_x=1} &= \text{det}(\bm{\mathcal{J}}_{xx,yy,xy})^{-1/3}\nonumber\\
    &\geq \frac{[(2+\gamma)(1-\gamma)(1+2\gamma)]^{1/3}}{3}\\
    \left.\bar{\sigma}^{2}_{xx,yy,xy}\right|_{\bar{\mu}_z=1} &\geq \frac{2^{1/3}(1-\gamma)}{3}
\end{align}
\end{subequations}\end{linenomath}
for molecules wobbling around $\mu_x$ and $\mu_z$ axes, respectively. Similarly, in the limiting case where $\mathcal{J}_{xy,xy}=\mathcal{J}_{yz,yz}=0$, the bound on the total measurement variance for $m_{xx}$, $m_{zz}$, and $m_{xz}$ [\cref{fig:2}(b)] is given by 
\begin{linenomath}\begin{subequations}
\label{eqn:xzBound}
\begin{align}
    \left.\bar{\sigma}^2_{xx,zz,xz}\right|_{\bar{\mu}_x=1} &= \text{det}(\bm{\mathcal{J}}_{xx,zz,xz})^{-1/3}\nonumber\\*
    &\geq \frac{\{c(1-\gamma)(1+2\gamma)[1+c+(2-c)\gamma]\}^{1/3}}{3c}\\
    \left.\bar{\sigma}^2_{xx,zz,xz}\right|_{\bar{\mu}_z=1}&\geq \frac{\{c(1-\gamma)(1+2\gamma)[1+c+(2c-1)\gamma]\}^{1/3}}{3c},
\end{align}
\end{subequations}\end{linenomath}
where $c$ represents the brightness of a $z$-oriented dipole normalized to that of an $x$- or $y$-oriented dipole [\cref{eqn:basis_normalization}]. A bound on the uncertainty of measuring all six second moments (using the full FI matrix $\bm{\mathcal{J}}$) can be found analytically by solving
\begin{equation}
\label{eqn:fullMbound}
    \nabla_{[\mathcal{J}_{12},\mathcal{J}_{13},\mathcal{J}_{23}]}f(\bm{\mathcal{J}})=\bm{0}
\end{equation} 
using the aforementioned constraints [\cref{eqn:FIconstraint_x1,eqn:FIconstraint_x2,eqn:FIconstraint_z}]. However, due to the complexity of the expression, we only show the numerical results here [\cref{fig:2}(c)]. Note that the classical precision bound [\cref{eqn:xyBound,eqn:xzBound,eqn:fullMbound}] lies above (i.e., predicts worse precision than) the quantum bound for measuring orientational second moments \cite{zhang2020quantum}. Achieving quantum-limited performance for measuring all second-order orientational moments simultaneously requires collective measurement \cite{Ragy2016,Bisketzi2019}, which is not usually possible in SMOLM since fluorescent probes and photon detection events are uncorrelated.

With this fundamental bound on measurement precision in hand, we now compare the sensitivity of various methods for measuring the orientation of in-focus molecules wobbling around the $\mu_x$ [\cref{fig:2}(i)] and $\mu_z$ axes [\cref{fig:2}(ii)]. Throughout the following analyses, we assume an imaging numerical aperture of $\text{NA}=1.4$ and a matched immersion RI of $n=1.515$, if not otherwise specified. Interestingly, the raPol standard PSF measures the in-plane second moments $m_{xx}$, $m_{yy}$, and $m_{zz}$ for molecules wobbling around either the $\mu_x$-axis [\cref{fig:2}(a)(i)] or the $\mu_z$ axis [\cref{fig:2}(a)(ii)] with precision close to the fundamental limit (within 11\% on average); thus raPol is nearly optimal for measuring the in-plane rotational dynamics of fluorescent molecules. The $xy$Pol, TS, and bisected PSFs all have worse precision due to the weak $B_{xy}$ basis images discussed in \cref{sec:model}.

In contrast, the family of standard PSFs has weak $B_{xz}$ and $B_{yz}$ basis images, and therefore, their precisions for measuring moments in the $xz$ plane are much worse compared to those of engineered PSFs [\cref{fig:2}(b)]. We also note that none of the PSFs we evaluate perform close to the best-possible measurement precision. Similar observations hold for measuring all second moments simultaneously [\cref{fig:2}(c)]; raPol and $xy$Pol have relatively poor precision, and all methods are significantly worse than the bound. 
\begin{figure}[ht!]
    \centering
    \includegraphics[width=8.8cm]{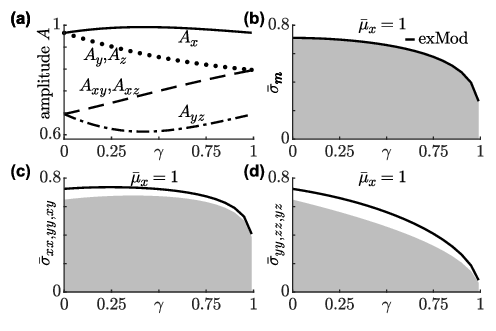}
    \caption{Excitation modulation (exMod) scheme [\cref{eqn:exPols}] that achieves the best-possible 3D orientation precision for a molecule wobbling symmetrically around the $\mu_x$ axis. (a) Normalized brightness of each pumping polarization $\{A_i,A_{ij}\}$ in the scheme. (b,c,d) Limit of estimation precision (b) $\bar{\sigma}_{\bm{m}}$ for all 3D second-order orientational moments $\bm{m}$; (c)~$\bar{\sigma}_{xx,yy,xy}$ for $m_{xx}$, $m_{yy}$, and $m_{xy}$; and (d) $\bar{\sigma}_{yy,zz,yz}$ for $m_{yy}$, $m_{zz}$, and $m_{yz}$ for one photon detected. For $N$ photons detected, scale the values by $1/\sqrt{N}$.}
    \label{fig:3}
\end{figure}

Although it is difficult to directly use these FI bounds to guide the design of $\bm{J}$ for optimal precision, it is possible to find sets of optimal excitation polarizations \cite{backer2016enhanced,Blanchard2020} that achieve the best-possible measurement precision by simply counting photons emitted in response to each polarization (\cref{fig:3}). For simplicity, we use 9 linearly polarized pumping fields $\{\bm{E}_{x0},$ $\bm{E}_{y0},$ $\bm{E}_{z0},$ $\bm{E}_{x0}\pm\bm{E}_{y0},$ $\bm{E}_{x0}\pm\bm{E}_{z0},$ $\bm{E}_{y0}\pm\bm{E}_{z0}\}$ (parallel to each second-order orientational moment) with 6 independent brightness scaling factors, i.e.,
\begin{linenomath}\begin{subequations} \label{eqn:exPols}
\begin{align}
    [\bm{E}_{x0},\bm{E}_{y0},\bm{E}_{z0}] &= \begin{bmatrix}1&0&0\\0&1&0\\0&0&1\end{bmatrix}\\
    \bm{E}_i &= A_i\bm{E}_{i0}\\
    \bm{E}_{ij,1} &= A_{ij}(\bm{E}_{i0}+\bm{E}_{j0})/\sqrt{2}\\
    \bm{E}_{ij,2} &= A_{ij}(\bm{E}_{i0}-\bm{E}_{j0})/\sqrt{2},
\end{align}
\end{subequations}\end{linenomath}
where $\{i,j\}\in\{x,y,z\}$ and $i\neq j$. It is simple to solve for a set of brightness scaling factors $A_i$ and $A_{ij}$ [\cref{fig:3}(a)] such that using this sequence of polarized excitation beams (\cref{eqn:exPols}) achieves the classical performance limit (\cref{sec:orientation}) for measuring the 3D orientation of a molecule wobbling symmetrically around the $\mu_x$ axis [\cref{fig:3}(b)]. However, this condition cannot satisfy the classical bound in measuring orientation and wobble within the $xy$-plane [\cref{fig:3}(c)] or $yz$-plane [\cref{fig:3}(d)] individually. Hence, there exists a trade-off between measuring optimally the in-plane moments versus measuring all 3D moments with maximum precision.

\section{Quantum limits for localizing fluorescent molecules}
\label{sec:localization}
In this section, we derive a minimum bound on the uncertainty of localizing a single molecule. Recently, researchers have adapted quantum estimation theory to develop an instrument-independent bound, i.e., the quantum Cram\'er-Rao bound (QCRB), on the sensitivity of estimating the position of fluorescent molecules \cite{tsang2015quantum,tsang2016quantum,lupo2016,Rehacek:17,ang2017quantum,tsang2019,prasad2019}. The quantum Fisher information (QFI) matrix $\bm{\mathcal{K}_r}$, which is the inverse of the QCRB, bounds the classical FI matrix $\bm{\mathcal{J}_r}$ such that $\bm{\mathcal{J}_r}^{-1}-\bm{\mathcal{K}_r}^{-1}$ is always positive semidefinite. In 2018, Backlund \emph{et al.}\ \cite{backlund2018} derived the QCRB associated with a scalar diffraction model which leverages the monopole approximation \cite{Chandler:19,Chandler2019,Chandler2020}. Here, we find the QCRB for estimating the position of dipole-like emitters using our vectorial forward imaging model [\cref{sec:model,eqn:model}]. 

\subsection{Quantum Fisher information for localizing rotationally fixed molecules}
We first develop a family of vectorial wavefunctions $\bm{\psi}_{\bm{\mu}}(u,v)$ to represent a photon collected from a dipole emitter with arbitrary orientation $\bm{\mu}=[\mu_x,\mu_y,\mu_z]^\intercal$, $\left\|\bm{\mu}\right\|=1$. The angular band-limited nature of dipole radiation collected by an imaging system \cite{Chandler:19} enables us to write the optical field from any dipole as a linear combination of three basis fields in each of two orthogonal polarization states [\cref{eqn:model}] that are paraxial in the BFP. Throughout this section, we use a subscript $(\cdot)_l$ to represent quantities associated with a dipole of molecular orientation $\mu_l$, while superscripts $(\cdot)^{(p)}$ represent quantities corresponding to a photon with polarization $p$.

Therefore, the $p^{\text{th}}$ polarization component of the wavefunction associated with any dipole emitter is given by
\begin{linenomath}
\begin{multline} \label{eqn:polarized_wavefunction}
    \psi_{\bm{\mu}}^{(p)}(u,v) = \frac{\mu_x\,g_x^{(p)}(u,v)+\mu_y\,g_y^{(p)}(u,v)+\mu_z\,g_z^{(p)}(u,v)}{\left(\mu_x^2+\mu_y^2+c\mu_z^2\right)^{1/2}} \\
    \times \exp\left[j\,\theta(u,v;x,y,z)\right]\\
    = \frac{\mu_x\,\check{g}_x^{(p)}(u,v)+\mu_y\,\check{g}_y^{(p)}(u,v)+\mu_z\,\check{g}_z^{(p)}(u,v)}{\left(\mu_x^2+\mu_y^2+c\mu_z^2\right)^{1/2}},
\end{multline}
\end{linenomath}
where $p=1$ represents $x$-polarized light, $p=2$ represents $y$-polarized light, and the phase delay $\theta(u, v;x,y,z)$ arises from the dipole's position [\cref{eqn:phase_delay}]. In addition, the constants $c$, $c_1$, $c_2$, and $\epsilon$ [\cref{eqn:basis_normalization}] normalize each polarization component such that
\begin{linenomath}\begin{subequations}
\begin{align}
    &\iint\left| \psi_{\bm{\mu}}^{(1)}(u,v) \right|^2\,du\,dv = c_1^2\nonumber\\*
    =&\ \frac{(15+6\epsilon+3\epsilon^2)\mu_x^2+(1-2\epsilon+\epsilon^2)\mu_y^2+(8-4-4\epsilon^2)\mu_z^2}{4(4+\epsilon+\epsilon^2)(\mu_x^2+\mu_y^2)+4(4-2\epsilon-2\epsilon^2)\mu_z^2}\\ 
    &\ \iint \left| \psi_{\bm{\mu}}^{(2)}(u,v) \right|^2\,du\,dv = c_2^2\nonumber\\*
    =&\frac{(1-2\epsilon+\epsilon^2)\mu_x^2+(15+6\epsilon+3\epsilon^2)\mu_y^2+(8-4-4\epsilon^2)\mu_z^2}{4(4+\epsilon+\epsilon^2)(\mu_x^2+\mu_y^2)+4(4-2\epsilon-2\epsilon^2)\mu_z^2}\\ 
    &c_1^2+c_2^2=1.
\end{align}
\end{subequations}\end{linenomath}

Computing the QFI of generalized photon states often involves complicated eigendecomposition to find symmetric logarithmic derivatives (SLDs), so we therefore develop an analytical QFI expression specifically for pure photon states corresponding to light emitted by rotationally fixed molecules. Assuming quasimonochromatic and paraxial photons within our imaging system (see \cref{sec:model}), we follow established methods to quantize the paraxial electromagnetic field \cite{Aiello2005,Hawton2007} and define a position-polarization eigenstate as 
\begin{equation}
    \ket{u,v,p} \equiv a^\dagger(u,v,p)\ket{0},
\end{equation} 
where the creation operator $a^\dagger(u,v,p)$ satisfies the commutation relation $[a(u,v,p),a^\dagger(u',v',p')] = \delta(u-u', v-v')\,\delta_{pp'}$ and $\delta(u-u',\allowbreak v-v')$ and $\delta_{pp'}$ are the Dirac and Kronecker delta functions, respectively. 

This definition makes it possible to build a single-photon state 
\begin{equation} \label{eqn:psi_ket}
    \ket{\psi_{\bm{\mu}}} \equiv c_1\ket{1} + c_2\ket{2}
\end{equation}
corresponding to any dipole emitter, where $\ket{1}$ represents an $x$-polarized photon and $\ket{2}$ represents a $y$-polarized photon. We define \cite{Aiello2005}
\begin{subequations}
\label{eqn:quantization}
\begin{align}
    \psi_{\bm{\mu}}^{(p)}(u,v) &\equiv \ip{u,v,p}{\psi_{\bm{\mu}}}, \text{ such that}\\
    \ket{p} &= c_p^{-1}\iint \psi_{\bm{\mu}}^{(p)}(u,v) \ket{u,v,p}\,du\,dv, \text{ and} \label{eqn:p_ket} \\
    \ip{u,v,p'}{p} &= c_p^{-1} \psi_{\bm{\mu}}^{(p)}(u,v)\,\delta_{pp'},\quad p,p'\in\{1,2\}.
\end{align}
\end{subequations}
This formulation leads to intuitive definitions for the partial derivative of $\ket{\psi_{\bm{\mu}}}$ with respect to parameter $i$, given by
\begin{align}
\label{eqn:quantizationDerivative}
     \frac{\partial}{\partial i} \ket{\psi_{\bm{\mu}}} = \sum_{p=1}^2 \iint \frac{\partial \psi_{\bm{\mu}}^{(p)}(u,v)}{\partial i} \ket{u,v,p} \,du\,dv \equiv \ket{\frac{\partial \psi_{\bm{\mu}}}{\partial i}},
\end{align}
and the inner product between two arbitrary states, written as
\begin{align}
\label{eqn:innerProduct}
    \ip{\psi_{\bm{\mu}_1}}{\psi_{\bm{\mu}_2}} = \sum_{p=1}^2\iint \psi_{\bm{\mu}_1}^{*(p)}(u,v)\,\psi_{\bm{\mu}_2}^{(p)}(u,v)\,du\,dv.
\end{align}

It is now helpful to write the photon states associated with $x$-, $y$-, and $z$-oriented molecules and their partial derivatives with respect to position $i\in\{x,y,z\}$ as
\begin{subequations}
\label{eqn:basisFieldKet}
\begin{align}
    \ket{\psi_l}&=\sum_{p=1}^2\iint \check{g}_l^{(p)}(u,v)
    \ket{u,v,p}\,du\,dv,\quad l\in\{x,y\}\\
    \ket{\psi_z}&=c^{-1/2}\sum_{p=1}^2\iint \check{g}_z^{(p)}(u,v)
    \ket{u,v,p}\,du\,dv\\
    \ket{\pdv{\psi_l}{i}}&=\sum_{p=1}^2\iint \pdv{\check{g}_l^{(p)}(u,v)}{i}\ket{u,v,p}\,du\,dv,\\
    \ket{\pdv{\psi_z}{i}}&=c^{-1/2}\sum_{p=1}^2\iint \pdv{\check{g}_z^{(p)}(u,v)}{i}\ket{u,v,p}\,du\,dv
\end{align}
\end{subequations}
Substituting \cref{eqn:polarized_wavefunction,eqn:basisFieldKet} into \cref{eqn:psi_ket,eqn:p_ket,eqn:quantizationDerivative}, we obtain
\begin{linenomath}\begin{subequations}
\begin{align}
    \ket{\psi_{\bm{\mu}}} &= \frac{\mu_x\ket{\psi_x}+\mu_y\ket{\psi_y}+c^{1/2}\mu_z\ket{\psi_z}}{\left(\mu_x^2+\mu_y^2+c\mu_z^2\right)^{1/2}} \\
    \ket{\pdv{\psi_{\bm{\mu}}}{i}} 
    &= \frac{\mu_x\ket{\partial\psi_x/\partial i}+\mu_y\ket{\partial\psi_y/\partial i}+c^{1/2}\mu_z\ket{\partial\psi_z/\partial i}}{\left(\mu_x^2+\mu_y^2+c\mu_z^2\right)^{1/2}}.
\end{align}
\end{subequations}\end{linenomath} 
We have now linked the single-photon state $\ket{\psi_{\bm{\mu}}}$ associated with a dipole of arbitrary orientation to the states corresponding to $x$-, $y$-, and $z$-oriented dipoles, and thus the classical basis fields $\check{g}_l^{(p)}$ [\cref{eqn:basis_field}], in a direct manner; their relative impact is controlled simply by the orientation $\bm{\mu}$ of the dipole.

The entries of the QFI matrix can be calculated using \cite{helstrom1976quantum}
\begin{equation}
    \mathcal{K}_{ii'} = 4\text{Re}\left[\ip{\pdv{\psi_{\bm{\mu}}}{i}}{\pdv{\psi_{\bm{\mu}}}{i'}}+\ip{\pdv{\psi_{\bm{\mu}}}{i}}{\psi_{\bm{\mu}}}\ip{\psi_{\bm{\mu}}}{\pdv{\psi_{\bm{\mu}}}{i'}}\right],
\end{equation}
where $\{i,i'\}\in\{x,y,z\}$. Therefore, we compute the inner products between 12 state vectors $\ket{\psi_l}$ and their derivatives $\ket{\partial \psi_l/\partial i}$, $\{i,l\}\in\{x,y,z\}$. Due to the complexity of the expression, we write the QFI as the weighted sum of 5 component matrices, i.e.,
\begin{equation}
    \bm{\mathcal{K}_r}=\sum_{i=1}^5\lambda_i\bm{\nu}_i\bm{\nu}_i^\intercal.
\end{equation}
The components are each the outer product of unit vectors $\bm{\nu}_i$, given by
\begin{linenomath}\begin{subequations}
\begin{align}
    \bm{\nu}_1=\ &[\mu_x,\mu_y,\mu_z]^\intercal\\
    \bm{\nu}_2=\ &[0,0,1]^\intercal\\
    \bm{\nu}_3=\ &(1-\mu_z^2)^{-1/2}[\mu_x,\mu_y,0]^\intercal\\
    \bm{\nu}_4=\ &(1-\mu_z^2)^{-1/2}[-\mu_y,\mu_x,0]^\intercal\\
    \bm{\nu}_5=\ &\bm{\nu}_{5*}\big /\left\|\bm{\nu}_{5*}\right\|_2\\
    \bm{\nu}_{5*} =\ & [(-2+2\epsilon^2)\mu_x\mu_z,(-2+2\epsilon^2)\mu_y\mu_z,\nonumber\\
    &(1+3\epsilon^2)\mu_z^2-(3+\epsilon^2)]^\intercal,
\end{align}
\end{subequations}\end{linenomath}
weighted by scalars $\lambda_i$, written as
\begin{linenomath}\begin{subequations}
\begin{align}
    \lambda_1&=\frac{8k^2\pi A}{15}(1-\epsilon)^2(2+4\epsilon+6\epsilon^2+3\epsilon^3)\\
    \lambda_2&=\frac{4k^2\pi A}{15}(8-5\epsilon^3-3\epsilon^5)(1-\mu_z^2)\\
    \lambda_3&=\frac{k^2\pi A}{15}(1-\epsilon)^2\left[(32+19\epsilon+6\epsilon^2+3\epsilon^3)\mu_z^2-15\epsilon(1+\epsilon)^2\right]\\
    \lambda_4&=\frac{k^2\pi A}{15}\left[32(1-\epsilon)-(13\epsilon+3\epsilon^3)(1-\epsilon^2) - 15\epsilon(1-\epsilon^2)^2\mu_z^2\right]\\
    \lambda_5 &= -\frac{k^2\pi^2 A^2}{4}(1-\epsilon^2)^2\left\|\bm{\nu}_{5*}\right\|_2^2,
\end{align}
\end{subequations}\end{linenomath} 
where scaling factor $A=3\left\{\pi(1-\epsilon)[(\epsilon+\epsilon^2)(1-3\mu_z^2)+4]\right\}^{-1}$ normalizes the amplitude of the wavefunction to represent one photon.

Although $\lambda_i$ and $\bm{\nu}_i$ are not the eigenvalues and eigenvectors of the QFI matrix $\bm{\mathcal{K}_r}$, each component varies symmetrically around the optical axis (the $\mu_z$ direction). The first term in the summation $\bm{\nu}_1\bm{\nu}_1^\intercal$ represents an FI contribution from the outer product of the molecular orientation $\bm{\mu}$, weighted by a factor $\lambda_1$ that is independent of orientation. The next three terms $\bm{\nu}_2\bm{\nu}_2^\intercal$, $\bm{\nu}_3\bm{\nu}_3^\intercal$, and $\bm{\nu}_4\bm{\nu}_4^\intercal$ represent FI components from the outer products of the axial, in-plane radial, and in-plane azimuthal molecular orientations. The different weighting factors $\lambda_2$, $\lambda_3$, and $\lambda_4$ indicate the contributions of these axial and lateral molecular orientation components on the localization information contained within the imaging system PSF. The final term $\bm{\nu}_5\bm{\nu}_5^\intercal$ arises from $\ip{\partial\psi_{\bm{\mu}}/\partial i}{\psi_{\bm{\mu}}}\ip{\psi_{\bm{\mu}}}{\partial\psi_{\bm{\mu}}/\partial i'}$. Unlike the first four terms, its FI depends on higher-order molecular orientational moments.
\begin{figure}[tb!]
    \centering
    \includegraphics[width=8.8cm]{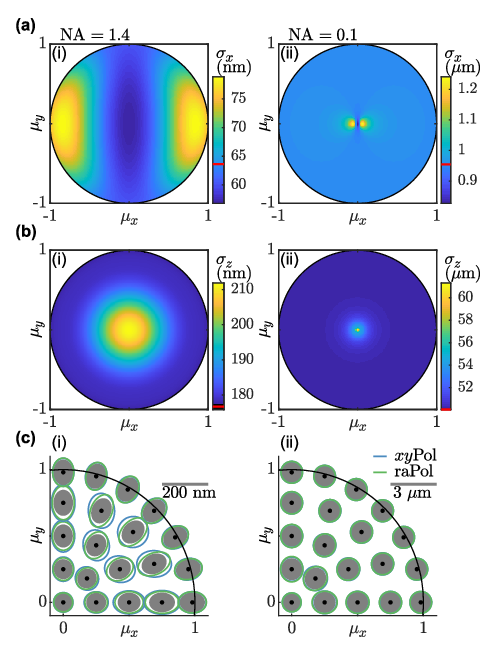}
    \caption{Quantum precision bounds $\sqrt{\text{QCRB}}$ for estimating the (a) lateral position $x$ and (b) axial position $z$ of rotationally-fixed dipole emitters with one photon detected using (i) high (1.4) and (ii) low (0.1) NA objective lenses. For $N$ photons detected, scale the values by $1/\sqrt{N}$. Red lines in the colorbars represent the $\sqrt{\text{QCRB}}$ derived from scalar diffraction theory \cite{backlund2018}. (c)~Error ellipses representing the lateral localization precision using polarized standard PSFs compared to the quantum bound represented by dark gray areas. Blue: $xy$Pol, green: raPol.}
    \label{fig:4}
\end{figure}

Here, we show a numerically computed quantum bound on the best-possible localization precision [$\sqrt{\text{QCRB}}$, \cref{fig:4}(a,b)(i)] attainable by any imaging system with $\text{NA}=1.4$ and $n=1.515$. Due to the toroidal emission pattern of fluorescent molecules, the image of an $x$-oriented molecule is elongated along the $x$ direction; therefore localization precision along the $x$ direction [\cref{fig:4}(a)(i)] is worse for $x$-oriented molecules compared to $y$-oriented molecules. The axial precision [\cref{fig:4}(b)(i)] is worse for $z$-oriented molecules compared to in-plane molecules. Interestingly, by using a polarized vectorial imaging model and considering the full dipole emission pattern, the quantum limits of localization precision are larger on average (10\% larger in the lateral direction, 3\% larger in the axial direction) compared to those derived using the scalar model \cite{backlund2018}. Intuitively, the quantum bound computed using the vectorial imaging model asymptotically approaches that of the scalar imaging model for small NAs for almost every molecular orientation [\cref{fig:4}(a,b)(ii)].
\begin{figure}[bt!]
    \centering
    \includegraphics[width=8.8cm]{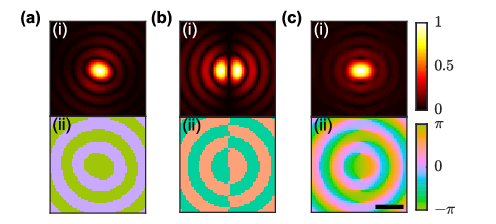}
    \caption{(i) Normalized amplitude and (ii) phase of the $x$-polarized optical field detected at the imaging plane using the standard PSF with $x$- and $y$-polarization separation ($xy$Pol) for a molecule with orientation (a) $\bm{\mu}=[1/\sqrt{2},1/\sqrt{2},0]^\intercal$, (b)~$\bm{\mu}=[0,0,1]^\intercal$, and (c)~$\bm{\mu}=[1/\sqrt{2},0,1/\sqrt{2}]^\intercal$. Scalebar: 500 nm.}
    \label{fig:5}
\end{figure}

We next compare the localization precision attainable using polarized standard PSFs to the quantum limit [\cref{fig:4}(c)(i)]; these PSFs are usually considered to be optimal for measuring the lateral position of in-focus molecules. Interestingly, we notice that although the standard PSFs can saturate the quantum bound for molecules that are parallel ($\mu_z=1$) or perpendicular ($\mu_z=0$) to the optical axis, their precisions for intermediate orientations are worse than the quantum bound. This performance variation arises from the phase content of the electric field in the image plane; for in-plane molecules, e.g., $\bm{\mu}=[1/\sqrt{2},1/\sqrt{2},0]^\intercal$ [\cref{fig:5}(a)] or molecules parallel to the optical axis [\cref{fig:5}(b)], the optical field at the image plane strictly contains only binary phase values [\cref{fig:5}(a,b)(ii)] as discussed in \cref{sec:model}. Therefore, when a camera captures an image, i.e., the squared magnitude of the optical field [\cref{fig:5}(a,b)(i)], no information is lost in the measurement. However, for an intermediate molecular orientation, e.g., $\bm{\mu}=[1/\sqrt{2},0,1/\sqrt{2}]^\intercal$, \emph{both} the magnitude [\cref{fig:5}(c)(i)] and phase patterns [\cref{fig:5}(c)(ii)] of the optical field contain variations useful for estimating a molecule's position. Phase information is ignored by a conventional camera, and the resulting localization precision is worse than the quantum bound. Further, since raPol reduces the phase complexity by always having a binary phase pattern in the azimuthally polarized channel, its localization precision is better than that of $xy$Pol for intermediate molecular orientations [\cref{fig:4}(c)(i)]. For imaging systems with smaller NA, both standard PSFs perform closely to the quantum bound [\cref{fig:4}(c)(ii)].

\subsection{Quantum Fisher information for localizing wobbling molecules}
We now numerically evaluate the QFI for localizing wobbling molecules. For simplicity, we only analyze molecules symmetrically wobbling around the $\mu_x$, $\mu_y$, or $\mu_z$ axes, i.e., $m_{xy}=m_{xz}=m_{yz}=0$ with orientational second moments $m_{ij}$ defined as in \cref{eqn:rotationSimpleModel}. The QFI for arbitrarily oriented molecules can be obtained following a similar procedure. The mixed-state photon density matrix is given by
\begin{equation}
    \label{eqn:wobble_density}
    \rho = \frac{m_{xx}\op{\psi_x}+m_{yy}\op{\psi_y}+cm_{zz}\op{\psi_z}}{m_{xx}+m_{yy}+cm_{zz}}.
\end{equation}
To compute the QFI $\bm{\mathcal{K}_r}$ associated with $\rho$, we need to compute the SLDs $\mathcal{L}_{i}$ associated with molecular position $[x,y,z]^\intercal$, that is
\begin{equation}
\label{eqn:QFI_SLD}
    \mathcal{K}_{il}=\frac{1}{2}\Re{\Tr \rho\left(\mathcal{L}_{i}\mathcal{L}_{l}+\mathcal{L}_{l}\mathcal{L}_{i}\right)}, \quad \{i,l\}\in\{x,y,z\},
\end{equation}
where the SLDs are defined implicitly by \cite{tsang2016quantum}
\begin{equation}
    \pdv{\rho}{i}=\frac{1}{2}(\mathcal{L}_i\rho+\rho\mathcal{L}_i), \quad i\in\{x,y,z\}.
\end{equation}
Partial derivatives of the density matrix with respect to molecular position $i\in\{x,y,z\}$ are given by
\begin{multline}
    \pdv{\rho}{i}=\frac{1}{m_{xx}+m_{yy}+cm_{zz}}\left[m_{xx}\left(\op{\psi_x}{\pdv{\psi_x}{i}}+\op{\pdv{\psi_x}{i}}{\psi_x}\right)\right.\\
    +m_{yy}\left(\op{\psi_y}{\pdv{\psi_y}{i}}+\op{\pdv{\psi_y}{i}}{\psi_y}\right)\\
    \left.+cm_{zz}\left(\op{\psi_z}{\pdv{\psi_z}{i}}+\op{\pdv{\psi_z}{i}}{\psi_z}\right)\right].
\end{multline}
From $\rho$ [\cref{eqn:wobble_density}], we compute the SLD $\mathcal{L}_i$ via eigendecomposition and obtain \cite{tsang2016quantum,backlund2018,zhang2020quantum}
\begin{linenomath} \begin{multline}
    \mathcal{L}_l =  2 \mathlarger{\sum_{i\in\{x,y,z\}}} \left(\op{\pdv{\psi_i}{z}}{\psi_i}+\op{\psi_i}{\pdv{\psi_i}{z}}\right) \\
    +\left(c{m_{zz}\ip{\pdv{\psi_l}{l}}{\psi_z}}-{m_{ll}\ip{\pdv{\psi_z}{l}}{\psi_l}}\right) \\
    \times\frac{2\left(\op{\psi_z}{\psi_l}-\op{\psi_l}{\psi_z}\right)}{m_{ll}+cm_{zz}}, \quad l\in\{x,y\},
\end{multline} \end{linenomath}
for localizing a molecule along the lateral direction and
\begin{equation}
    \mathcal{L}_z = 2 \mathlarger{\sum_{i\in\{x,y,z\}}} {\op{\pdv{\psi_i}{z}}{\psi_i}+\op{\psi_i}{\pdv{\psi_i}{z}}}
\end{equation}
along the axial direction. Evaluating the QFI for localizing an isotropic emitter [$\gamma=0$ in \cref{eqn:rotationSimpleModel}], we obtain the best-possible (quantum) localization precision as
\begin{linenomath}
\begin{subequations}
\begin{align}
    \sigma_{x} &= \frac{2}{k}\left[\frac{3(\epsilon^2 + \epsilon - 8)}{9\epsilon^6 + 18\epsilon^5 - 17\epsilon^4 - 52\epsilon^3 + 63\epsilon^2 + 98\epsilon - 119}\right]^{1/2}\\
    \sigma_{z} &= \frac{2}{k} \left[\frac{3(\epsilon^3 + 3\epsilon^2 + 6\epsilon + 8)}{(\epsilon - 1)^2(7\epsilon^3 + 15\epsilon^2 + 21\epsilon + 29)}\right]^{1/2}.
\end{align}
\end{subequations}
\end{linenomath}
\begin{figure}[tb!]
    \centering
    \includegraphics[width=8.8cm]{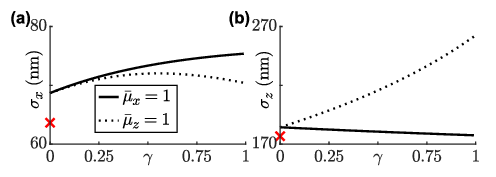}
    \caption{Quantum limit of precision $\sqrt{\text{QCRB}}$ for estimating the (a)~lateral and (b) axial position of molecules symmetrically wobbling around the $\mu_x$ (solid line) and $\mu_z$ (dotted line) axes. The precision is calculated assuming one photon is detected; for $N$ photons detected, scale the values by $1/\sqrt{N}$. Red crosses represent the $\sqrt{\text{QCRB}}$ derived based on a scalar diffraction model \cite{backlund2018}.}
    \label{fig:6}
\end{figure}

The QFIs for wobbling molecules are shown in \cref{fig:6}. Similar to the case of fixed molecules, the lateral and axial quantum precision bounds for isotropic emitters are 8\% and 4\% larger, respectively, than those derived using the scalar emission model \cite{backlund2018}. Therefore, QFIs derived based on scalar diffraction theory are overly optimistic on predicting the best-possible localization precision attainable for dipole-like emitters. 

\section{Discussion and Conclusion}
\label{sec:discussion}
In this work, we present a unified mathematical framework to compute the best-possible precision of measuring the orientation and position of dipole-like emitters independent of any specific instrument or microscope. We quantify the fundamental uncertainty of measuring the average orientation and wobble of a fluorescent molecule using the standard generalized variance (SGV) derived from the Cram\'er-Rao bound (CRB) and Fisher information (FI) matrix. Invoking the non-negativity of photon counting, we derive a classical precision bound that reveals that an optimal method for measuring the 2D $xy$-plane moments, e.g., the radially and azimuthally polarized standard PSF, cannot be optimal for measuring all orientational second moments in 3D simultaneously. Thus, it is impossible the design a fixed instrument that achieves the maximum sensitivity limit for measuring \emph{all possible} rotational motions.

Further, we extended existing quantum estimation theory on localizing point-like emitters to include the vectorial emission behavior of fluorescent molecules. Interestingly, we find that the best-possible localization precision for isotropic emitters is slightly larger, i.e., worse, than that derived using the scalar emission model \cite{backlund2018} when a high NA objective lens is used. These differences in predicted localization uncertainty are further evidence that the monopole and dipole emission models must be chosen with care \cite{Chandler:19,Chandler2019,Chandler2020}.

In this series's next paper \cite{location-orientation_2}, we compare the fundamental classical and quantum limits of orientation and localization precision to the performance of various SMOLM techniques; for several realistic sample configurations, we evaluate overall orientation-localization precision by combining the best-possible 2D/3D orientation measurement precision and 2D/3D localization precision into a single metric. Taken as a whole, our analysis suggests a new way forward for improving orientation and localization sensitivity in SMOLM and coping with a limited photon budget; maximum performance may be achieved by abandoning static instruments and measurement protocols and instead designing adaptive imaging systems that are optimized for specific measurement tasks. For example, one may consider recent innovations in nanometer-resolution SMLM, namely MINFLUX \cite{Balzarotti2017,Gwosch2020}, as adaptively changing the instrument PSF to localize molecules within a specific targeted region instead of an entire 3D volume simultaneously. Such an extension to orientation localization microscopy remains an exciting direction to be explored in the future.

\section*{Funding}
National Science Foundation (NSF) (1653777).

\section*{Acknowledgments}
We thank Miguel A. Alonso and Animesh Datta for their helpful insights on a preprint version of this paper and Jin Lu, Tianben Ding, Tingting Wu, and Hesam Mazidi for fruitful discussions.

\section*{Disclosures}
The authors declare no conflicts of interest.

\bibliography{references}

\end{document}